\documentclass[preprint,5p,times,twocolumn]{elsarticle}
\usepackage{upgreek}
\usepackage{amsmath}

\usepackage{amssymb}
\usepackage{units}
\usepackage{amsthm}

\usepackage{lineno}

\newcommand*\patchAmsMathEnvironmentForLineno[1]{%
	\expandafter\let\csname old#1\expandafter\endcsname\csname #1\endcsname
	\expandafter\let\csname oldend#1\expandafter\endcsname\csname end#1\endcsname
	\renewenvironment{#1}%
	{\linenomath\csname old#1\endcsname}%
	{\csname oldend#1\endcsname\endlinenomath}}%
\newcommand*\patchBothAmsMathEnvironmentsForLineno[1]{%
	\patchAmsMathEnvironmentForLineno{#1}%
	\patchAmsMathEnvironmentForLineno{#1*}}%
\AtBeginDocument{%
	\patchBothAmsMathEnvironmentsForLineno{equation}%
	\patchBothAmsMathEnvironmentsForLineno{align}%
	\patchBothAmsMathEnvironmentsForLineno{flalign}%
	\patchBothAmsMathEnvironmentsForLineno{alignat}%
	\patchBothAmsMathEnvironmentsForLineno{gather}%
	\patchBothAmsMathEnvironmentsForLineno{multline}%
}

\usepackage[utf8]{inputenc} 

\usepackage{graphicx}
\usepackage{subcaption}

\journal{Nuclear Instruments and Methods in Physics Research A}

\begin{document}

\begin{frontmatter}



\title{Seeding of the Self-Modulation in a Long Proton Bunch by Charge Cancellation with a Short Electron Bunch}

\author[MPP,TUM]{Mathias Hüther}
\author[MPP,CERN]{Patric Muggli}
\address[MPP]{Max Planck Institute for Physics, Munich, Germany}
\address[TUM]{Technical University of Munich, Garching b. München, Germany}
\address[CERN]{CERN, Geneva, Switzerland}

\begin{abstract}
In plasma wakefield accelerators (e.g. AWAKE) the proton bunch self-modulation is seeded by the ionization front of a high-power laser pulse ionizing a vapour and by the resulting steep edge of the driving bunch profile inside the created plasma.\\
In this paper, we present calculations in 2D linear theory for a concept of a different self-modulation seeding mechanism based on electron injection. The whole proton bunch propagates through a preformed plasma and the effective beam current is modulated by the external injection of a short electron bunch at the centre of the proton beam. The resulting sharp edge in the effective beam current in the trailing part of the proton bunch is driving large wakefields that can lead to a growth of the seeded self-modulation (SSM).\\
Furthermore, we discuss the feasibility for applications in AWAKE Run 2.

\end{abstract}


\end{frontmatter}

\section{Introduction}
\label{sec:intro}
\begin{figure}[htb!]
	\begin{subfigure}[b]{\textwidth}
		\includegraphics[width=0.45\linewidth]{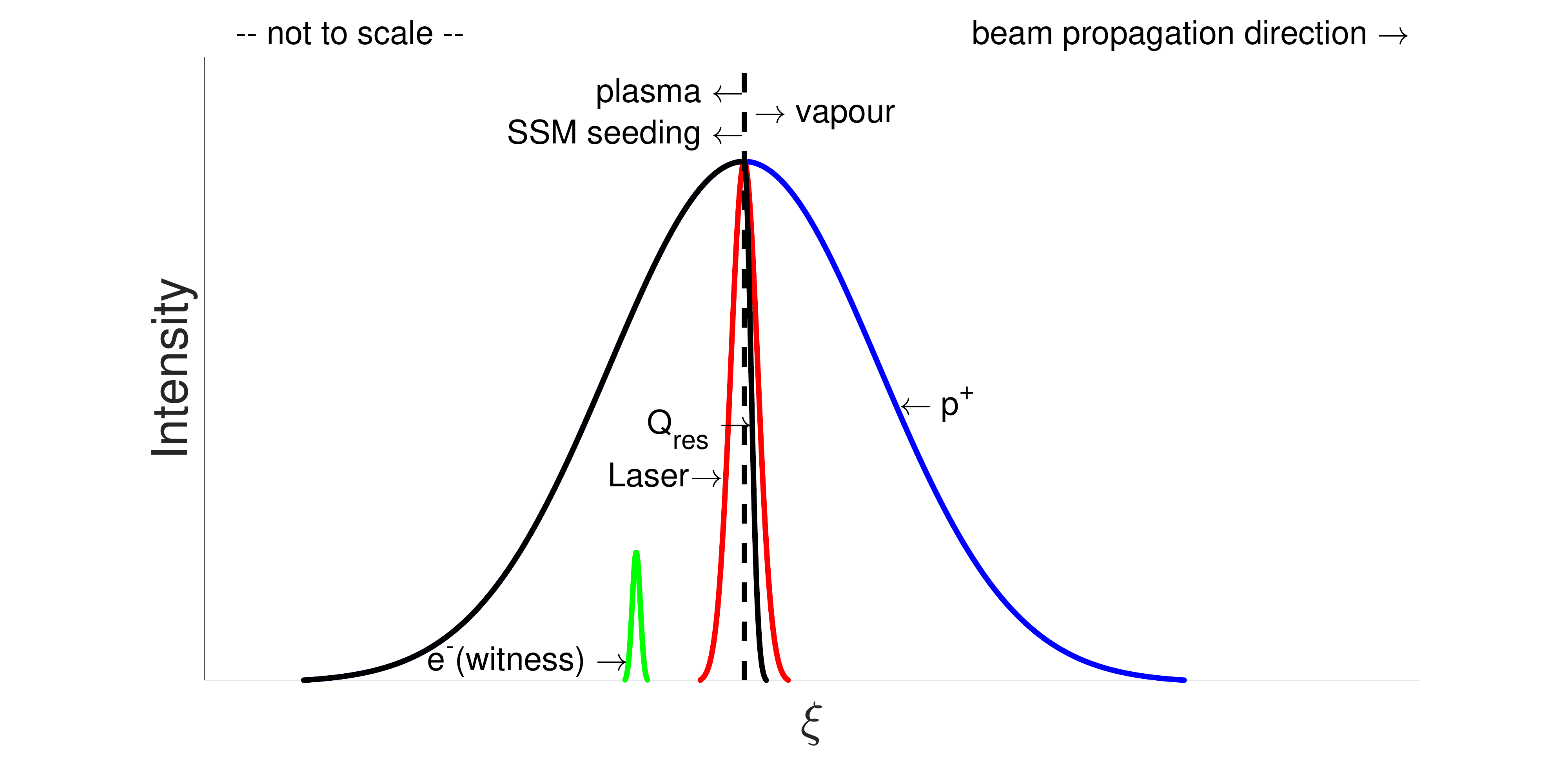}
		\caption{~~~~~~~~~~~~~~~~~~~~~~~~~~~~~}
		\label{fig:Run1_SeedingScheme}
	\end{subfigure}
	\begin{subfigure}[b]{\textwidth}
		\includegraphics[width=0.45\linewidth]{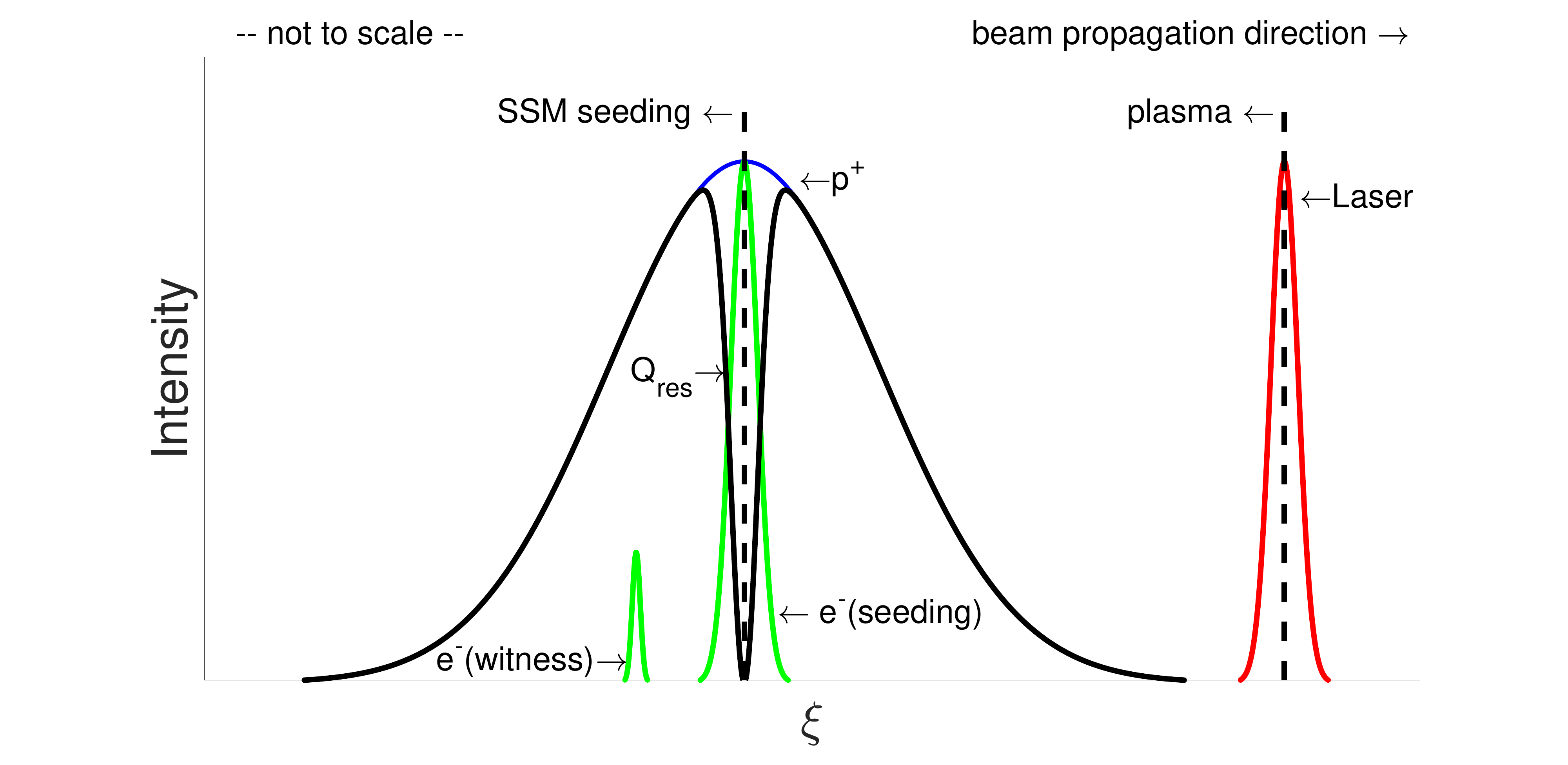}
		\caption{~~~~~~~~~~~~~~~~~~~~~~~~~~~~~}
		\label{fig:Run2_SeedingScheme}
	\end{subfigure}
	\begin{subfigure}[b]{\textwidth}
		\includegraphics[width=0.45\linewidth]{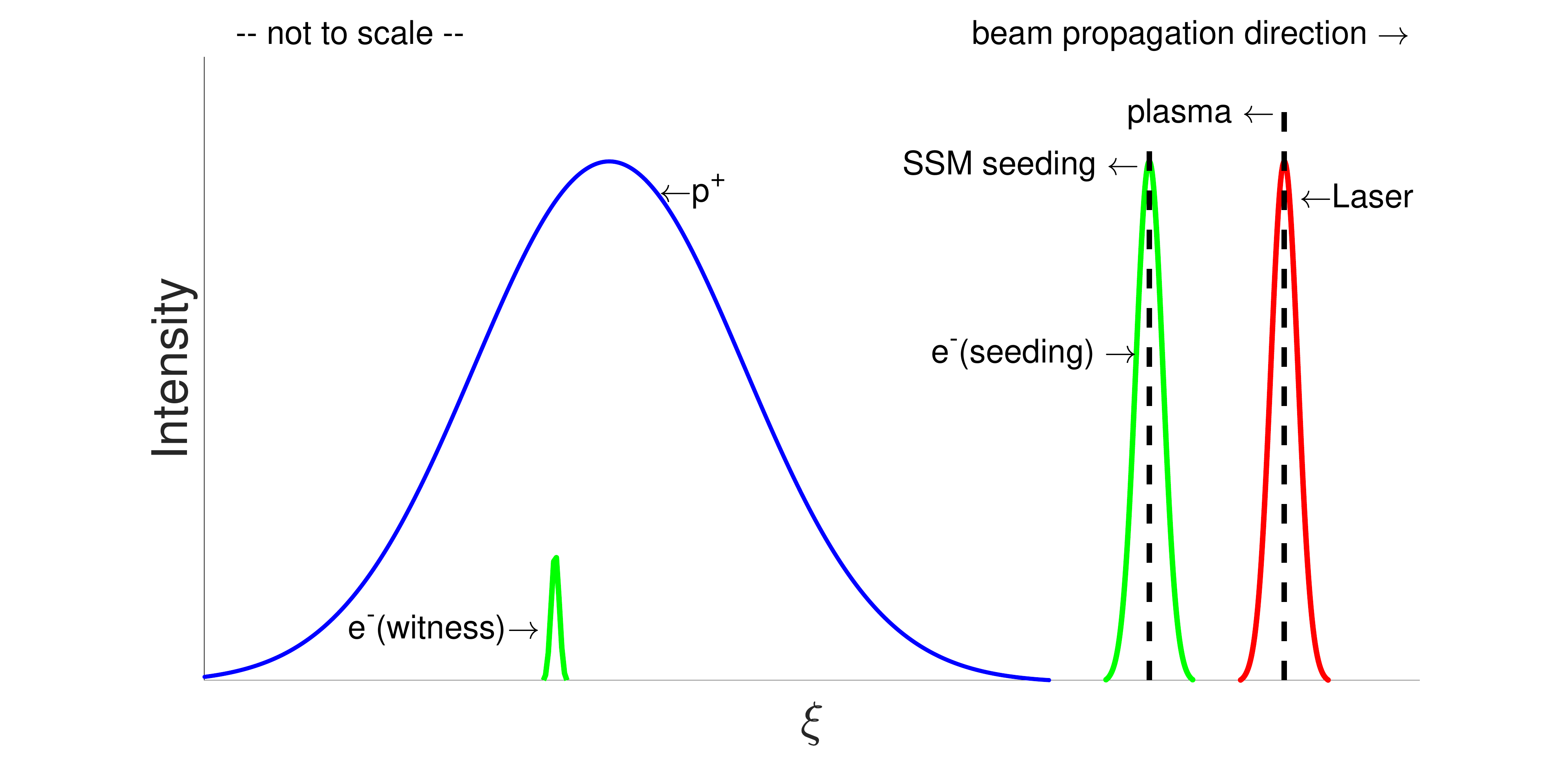}
		\caption{~~~~~~~~~~~~~~~~~~~~~~~~~~~~~}
		\label{fig:ElectronsAhead_SeedingScheme}
	\end{subfigure}
	\caption{Self-modulation seeding and electron injection schemes. (a) Current seeding concept with wakefields seeded by a vapour-plasma transition front created by a high-power laser pulse in the centre of the proton beam. (b) Novel seeding scheme with an electron bunch injected in the centre of the long proton bunch and resulting in a dip in the effective charge density. The laser pulse is ionizing the vapour ahead of the proton bunch. (c) Laser pulse and electron bunch ahead of the proton bunch. Seeding of SM process due to wakefields driven by the electron bunch.}
	\label{fig:SeedingSchemes}
\end{figure}

The proton bunch self-modulation in plasma wakefield accelerators (e.g. the Advanced Wakefield Experiment (AWAKE) located at CERN  \cite{muggli2013physics, gschwendtner2016awake}) is usually seeded by using a high-power laser pulse co-propagating in the centre of the proton bunch (see Fig.\,\ref{fig:Run1_SeedingScheme}). The laser ionizes a vapour and creates a plasma, resulting in a sharp relativistic ionization front separating vapour and plasma. Due to this ionization front, the proton bunch self-modulation is seeded and consequently growing within a distance of a few meters before reaching its saturation level \cite{PhysRevLett.112.194801,caldwell2016path}. The long proton bunch is split into a train of micro-bunches with a period on the order of the plasma wavelength \cite{kumar2010self}. This modulation is caused by the self-modulation instability (SMI), a transverse beam-plasma instability. A witness electron bunch can thus deterministically be injected into one of the buckets between the micro-bunches and accelerated.\\
The seeding concept presented in this paper gives another approach to reach this goal: A high-power laser pulse is ionizing the plasma some time ahead of the proton bunch, so that the proton bunch is propagating through a preformed plasma (see Fig.\,\ref{fig:Run2_SeedingScheme}). Hence, there is no seeding from the ionization process. A short electron bunch is injected at the centre or somewhere along the long proton bunch. The resulting local charge cancellation leads to a distinct drop in the effective charge density distribution - and is therefore driving wakefields in the rear part of the proton bunch. The front of the bunch is still long with a smoothly increasing density and thus drives much lower amplitude wakefields. Thanks to the seeding process \cite{lotov2013natural}, the wakefield phase and amplitude is determined and weakly dependent on initial beam parameters \cite{savard2017effect}. The witness electrons to be accelerated are again injected a number of buckets behind the seeding, as in the usual injection scheme. An overview of the different self-modulation seeding approaches using a short electron bunch is given in Fig.\,\ref{fig:SeedingSchemes}. A variation of the approach described in this paper, where a seed electron bunch is put ahead of the proton bunch, is shown in Fig.\,\ref{fig:ElectronsAhead_SeedingScheme} and will be explained later.\\
AWAKE is the world’s first proton-driven plasma wakefield accelerator aiming for acceleration of externally injected electrons in gradients up to the GV/m scale. It uses a 8-12 cm long proton bunch from CERN’s Super Proton Synchotron (SPS) that propagates through a 10 m long Rubidium plasma (with a density of $(1-10)\cdot10^{14}\,$cm$^{-3}$), induced by a high-power, short laser pulse (duration $\tau =  120\,$fs, energy $E = 450\,$mJ).

\section{A Simplified Model in Linear Theory}
\begin{figure}[htb!]
	\includegraphics[width=\linewidth]{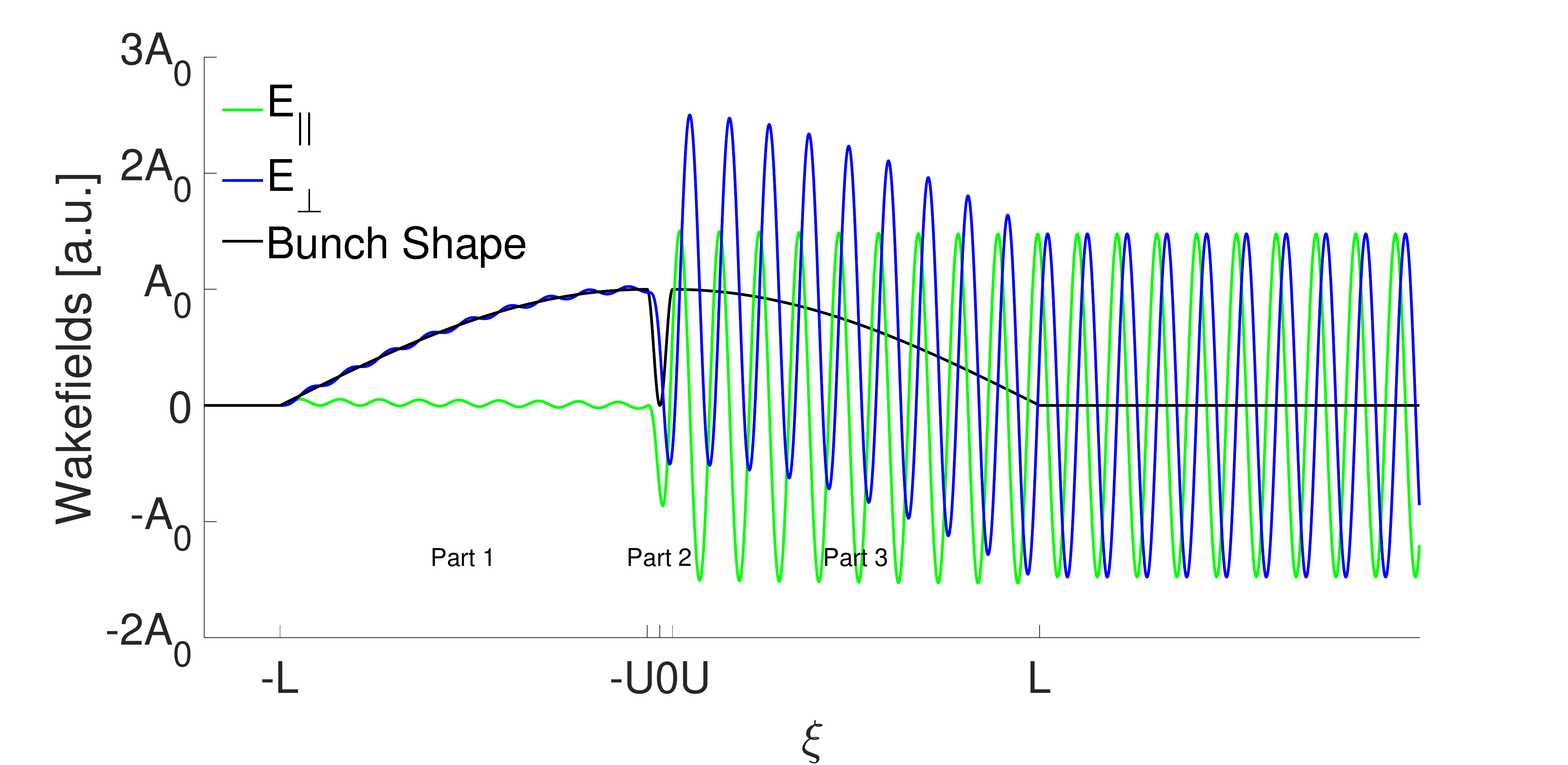}
	\caption{Parallel and perpendicular wakefields $E_\parallel(\xi)$ and $E_{\perp}(\xi)$  for a bunch shape with $\sigma_{z,p} =$ 12\,cm, $\sigma_{z,e} =$ 4\,mm, $\frac{U}{L} = \frac{1}{30}$, $\frac{U}{\lambda_{pe}} = 1.6$, peak currents $\mid Q_{p^+, peak}\mid = \mid Q_{e^-, peak} \mid$.}
	\label{fig: AWAKE-like case}
\end{figure}

For a more quantitative understanding of the approach towards seeding, we introduce the following simplified model: The proton and electron bunch longitudinal density distribution is described by a $cos$-profile, whereas their radial profiles are Gaussian. We use 2D-linear plasma wakefield theory \cite{chen1987plasma} to describe the seeding effect by calculating the initial wakefields.
\begin{figure}[htb!]
	\centering
	\includegraphics[width=0.72\linewidth]{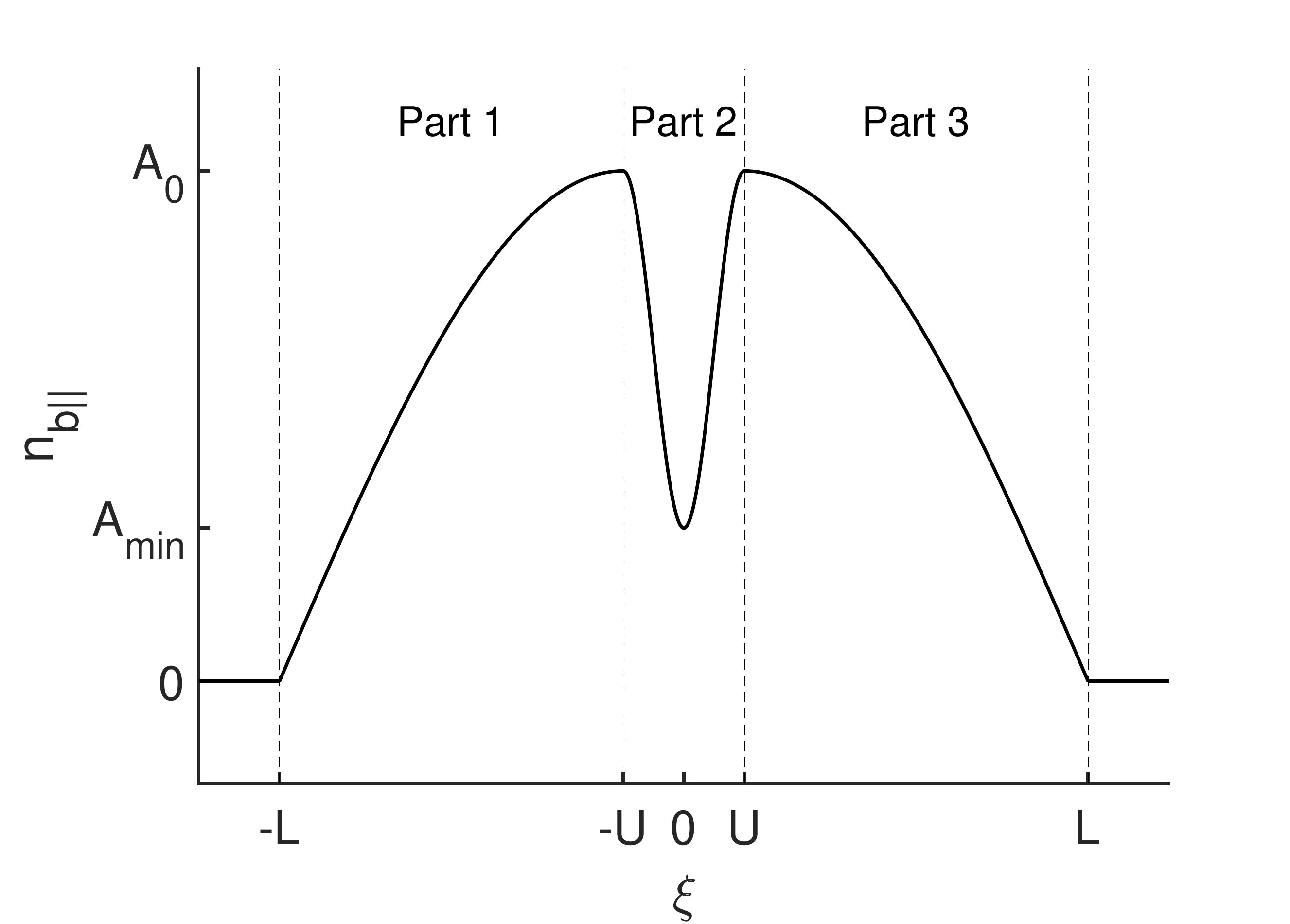}
	\caption{$\cos$-approximation of the longitudinal proton bunch density $n_{b\parallel}$ as described by Eq.\,\ref{eq:cos-approximation}.}
	\label{fig: bunch distribution}
\end{figure}
The total (proton and electron) longitudinal charge distribution $n_{b\parallel}$ (see Fig.\,\ref{fig: bunch distribution}) is given by
\begin{equation}
n_{b\parallel} (\xi') =
\begin{cases}
A_0 \cdot \sin\left(\frac{\pi \cdot (\xi' + L)}{2(L-U)}\right) &\xi\in [-L,-U]\\
\frac{A_{min} - A_0}{2} \cdot \cos \left(\frac{\pi\xi'}{U}\right) + \frac{A_{min} + A_0}{2} &\xi \in [-U,U]\\
A_0 \cdot \cos\left(\frac{\pi \cdot (\xi' - U)}{2(L-U)}\right) &\xi \in [U,L]\\
0  &~\text{elsewhere},
\end{cases}
\label{eq:cos-approximation}
\end{equation}
where $\xi = ct-z$ is the spatial coordinate in the co-moving bunch frame, $2L$ the longitudinal size of the proton bunch and $2U$ the longitudinal size of the electron bunch. Outside $[-L,L]$ the bunch charge density is assumed to be zero. The amplitude of the proton charge density is $A_0$ and the amplitude of the electron charge density is $A_0-A_{min}$ . The absolute number of charges in one bunch for our model is given by $N = 8 A_0 L n_{b} \sigma_{r,p}^2 \cdot 0.683$. For comparison in case of a Gaussian longitudinal bunch profile $N$ is given by $N = (2\pi)^{\frac{3}{2}} n_b\sigma_ {z,p} \sigma_{r,p}^2$.
According to \cite{chen1987plasma} the parallel and perpendicular wakefields $W_z(\xi, r)$ and $W_r(\xi, r)$ are given by
\begin{equation}
W_z(\xi, r) = \frac{e}{\epsilon_0} ~~\int_{-\infty}^{\xi} n_{b\parallel}(\xi') \cos\left(k_{pe}(\xi-\xi')\right) d\xi' \cdot ~R(r)
\label{eq: parallel wakefield}
\end{equation}
and
\begin{equation}
W_r(\xi, r) = \frac{e}{k_{pe} \epsilon_0} \displaystyle \int_{-\infty}^{\xi} n_{b\parallel}(\xi') \sin\left(k_{pe}(\xi-\xi')\right) d\xi'\cdot \frac{dR(r)}{dr},
\label{eq: perpendicular wakefield}
\end{equation}
where $e$ is the elementary charge, $\epsilon_0$ the vacuum permittivity and $k_{pe} = \omega_{pe}/v_b \approx \omega_{pe}/c$ the plasma wave number 
in case of a relativistic particle bunch with bunch velocity $v_b \approx c$. The plasma electron angular frequency $\omega_{pe}$ is
\begin{equation}
\omega_{pe} = \sqrt{\frac{n_{pe} e^2}{\epsilon_0 m_e}},
\end{equation}
where $m_e$ is the electron mass and $n_{pe}$ the plasma charge density. 
The transverse component $R(r)$ in Eqn.\,\ref{eq: parallel wakefield} and \ref{eq: perpendicular wakefield} is given by
\begin{align}
R(r) = k_{pe}^2~K_0(k_{pe}r)&\int_{0}^{r}r' n_{b\perp}(r')~I_0(k_{pe}r')~dr'\nonumber\\
 + k_{pe}^2~I_0(k_{pe}r)&\int_{r}^{\infty}r' n_{b\perp}(r')~K_0(k_{pe}r')~dr',
\end{align}
with $I_0$ and $K_0$ the modified Bessel functions of the first and second kind, respectively. Assuming the radial profile and the radial size of the electron and proton bunch are the same, $R(r)$ and $\frac{dR}{dr}$ are just multiplying factors. Considering different radii is trivial.\\
Piecewise integration of the separate bunch parts of Eqs.\,\ref{eq: parallel wakefield} and \ref{eq: perpendicular wakefield} along $\xi$ (see Fig.\,\ref{fig: bunch distribution}) gives the contribution of each single part to the resulting wakefields.\\
The wakefields driven by the leading part of the proton bunch $(-L < \xi < -U)$ are described by
\begin{equation}
\begin{footnotesize}
E_{1,\parallel~\,}(\xi) =
\begin{cases}
0\\
\frac{A_0}{2k_{pe}} \left[-(C_+ + C_-) \left[\cos\left(\frac{\pi(\xi+L)}{2(L-U)}\right) + \cos(k_{pe}(\xi+L)) \right]\right]\\
\frac{A_0}{2k_{pe}} \left[-(C_+ - C_-) \sin(k_{pe}(\xi+U)) + (C_+ +  C_-)\cos(k_{pe}(\xi+L))\right]\\
\end{cases}
\end{footnotesize}
\label{eq: E1par}
\end{equation}
and
\begin{equation}
\begin{footnotesize}
E_{1,\perp~}(\xi) =
\begin{cases}
0 \\
\frac{A_0}{2} \left[(C_+ - C_-)\sin\left(\frac{\pi(\xi+L)}{2(L-U)}\right) + (C_+ + C_-) \sin\left(k_{pe}(\xi+L)\right)\right] \\
\frac{A_0}{2}\left[(C_+ - C_-)\cos(k_{pe}(\xi + U)) + (C_+ + C_-) \sin(k_{pe}(\xi + L))\right],\\
\end{cases}
\end{footnotesize}
\end{equation}
where the first term is valid for $\xi < -L$, the second term for -$L \le \xi \le -U$ and the third term for $\xi > -U$. The parallel and perpendicular wakefields in the part where the electrons are injected and therefore cause a drop in the effective charge density are given by
\begin{equation}
\begin{footnotesize}
E_{2,\parallel}(\xi) = 
\begin{cases}
0 \\
\frac{A_1}{2k_{pe}}\left[(D_+ + D_-)\sin\left(\frac{\pi\xi}{U}\right) - (D_+ - D_-) \sin(k_{pe}(\xi+U))\right] \\~~~+ \frac{A_2}{k_{pe}^2}\sin(k_{pe}(\xi+U))\\
\frac{A_1}{2k_{pe}}\left[(D_+ - D_-)\sin(k_{pe}(\xi-U)) - (D_+ - D_-) \sin(k_{pe}(\xi+U))\right] \\~~~- \frac{A_2}{k_{pe}^2}\left[\sin(k_{pe}(\xi-U)) - \sin(k_{pe}(\xi+U))\right]\\
\end{cases}
\end{footnotesize}
\end{equation}
and
\begin{equation}
\begin{footnotesize}
E_{2,\perp}(\xi) = 
\begin{cases}
0 \\
\frac{A_1}{2} \left[(D_+ + D_-)\cos\left(\frac{\pi\xi}{U}\right) - (D_+ - D_-) \cos(k_{pe}(\xi + U))\right] \\~~~+ \frac{A_2}{k_{pe}}\left[1-\cos(k_{pe}(\xi+U))\right]\\
\frac{A_1}{2} \left[(-D_+ + D_-)\cos(k_{pe}(\xi-U)) + (D_+ - D_-) \cos(k_{pe}(\xi + U))\right] \\~~~- \frac{A_2}{k_{pe}}\left[\cos(k_{pe}(\xi+U)) - \cos(k_{pe}(\xi-U))\right].\\
\end{cases}
\end{footnotesize}
\end{equation}
Here, the first term is valid for $\xi < -U$, the second for $-U \le \xi \le U$ and the third for $\xi > U$.\\
The wakefields caused by bunch section between $U$ and $L$ are described by
\begin{equation}
\begin{footnotesize}
E_{3, \parallel~}(\xi) =
\begin{cases}
0 \\
\frac{A_0}{2k_{pe}}\left[(C_+ + C_-)\sin\left(\frac{\pi(\xi-U)}{2(L-U)}\right) + (C_+ - C_-)\sin(k_{pe}(\xi-U))\right] \\
\frac{A_0}{2k_{pe}}\left[(C_+ + C_-)\cos(k_{pe}(\xi-L)) + (C_+ - C_-)\sin(k_{pe}(\xi-U))\right]\\
\end{cases}
\end{footnotesize}
\end{equation}
and
\begin{equation}
\begin{footnotesize}
E_{3, \perp~}(\xi) =
\begin{cases}
0\\
\frac{A_0}{2} \left[(C_+ - C_-)\cos\left(\frac{\pi(\xi-U)}{2(L-U)}\right) + (-C_+ + C_-)\cos(k_{pe}(\xi-U))\right] \\
\frac{A_0}{2} \left[(C_+ + C_-)\sin(k_{pe}(\xi-L)) + (-C_+ + C_-)\cos(k_{pe}(\xi-U))\right].\\
\end{cases}
\end{footnotesize}
\label{eq: E3perp}
\end{equation}
For this case, the first line is valid for $\xi < U$, the second for the interval $ U \le \xi \le L$ and the third line for $\xi > L$.\\
The constants in Eqn.\,\ref{eq: E1par} to \ref{eq: E3perp} are given by $A_1 = \frac{A_{min}-A_0}{2}$, $A_2 = \frac{A_{min}+A_0}{2}$, $C\pm =  \frac{1}{\frac{\pi}{2(L-U)}\pm k_{pe}}$ and $D_\pm =  \frac{1}{\frac{\pi}{U}\pm k_{pe}}$.\\
Making use of the superposition theorem of waves in linear theory, the final wakefield distribution can be determined by adding the different contributions of the bunch segments in every region. Hence, the resulting wakefields are described by
\begin{equation}
E(\xi) =
\begin{cases}
0   &~\xi < -L\\
E_1(\xi)\big|_{\xi \in \left[-L;-U\right]} &\xi \in \left[-L;-U\right]\\
E_1(\xi)\big|_{\xi \in \left[-U;U\right]} + E_2(\xi)\big|_{\xi \in \left[-U;U\right]}  &\xi \in \left[-U;U\right]\\
E_1(\xi)\big|_{\xi \in \left[U;L\right]} + E_2(\xi)\big|_{\xi \in \left[U;L\right]} +E_3(\xi)\big|_{\xi \in \left[U;L\right]}  &\xi \in \left[U;L\right]\\
E_1(\xi)\big|_{\xi > L} + E_2(\xi)\big|_{\xi > L} +E_3(\xi)\big|_{\xi > L}  &\xi > L.\\
\end{cases}
\label{eq: parallel and perpendicular wakefields}
\end{equation}
Fig.\,\ref{fig: AWAKE-like case} shows the longitudinal and perpendicular wakefields (Eq.~\ref{eq: parallel and perpendicular wakefields}) for a ratio $\frac{U}{L} = \frac{1}{30}$ and $\frac{U}{\lambda_{pe}} = 1.6$. The front of the proton bunch ($\xi < -U$) propagates in a preformed plasma where its length is very long when compared to the plasma wavelength $\lambda_{pe}$ and its profile is smooth. It is therefore not effective at driving wakefields, but still drives a low amplitude fields that could lead to development of the SMI process over long plasma distances. The adiabatic response of the plasma generates a globally focussing force by charge neutralisation that, upon propagation, can lead to an increase in bunch density and thus also of the wakefield response. This effect could also lead to SMI growth (as opposed to SSM) over long plasma distances. From the position of the electron bunch ($\xi > U$) and all along the second half of the proton bunch, much larger wakefields are driven, which subsequently provide the strong seed for the SM process.\\
For bunch parameters similar to AWAKE ($L=$ 6\,cm, $U=$ 2\,mm, $\frac{U}{L} = \frac{1}{30}$, $\frac{U}{\lambda_{pe}} = 1.6$, peak currents $\mid Q_{p^+, peak}\mid = \mid Q_{e^-, peak} \mid$ (see Fig.\,\ref{fig: AWAKE-like case}) and $k_{pe} = 5000\,\text{m}^{-1}$, $n_p = 7\cdot 10^{14}\,\text{cm}^{-3}$, $\sigma_r = 200\,\upmu\text{m}$) the wakefields in the trailing part of the bunch could reach values up to the GV/m scale.

\begin{figure}[htb!]
	\includegraphics[width=\linewidth]{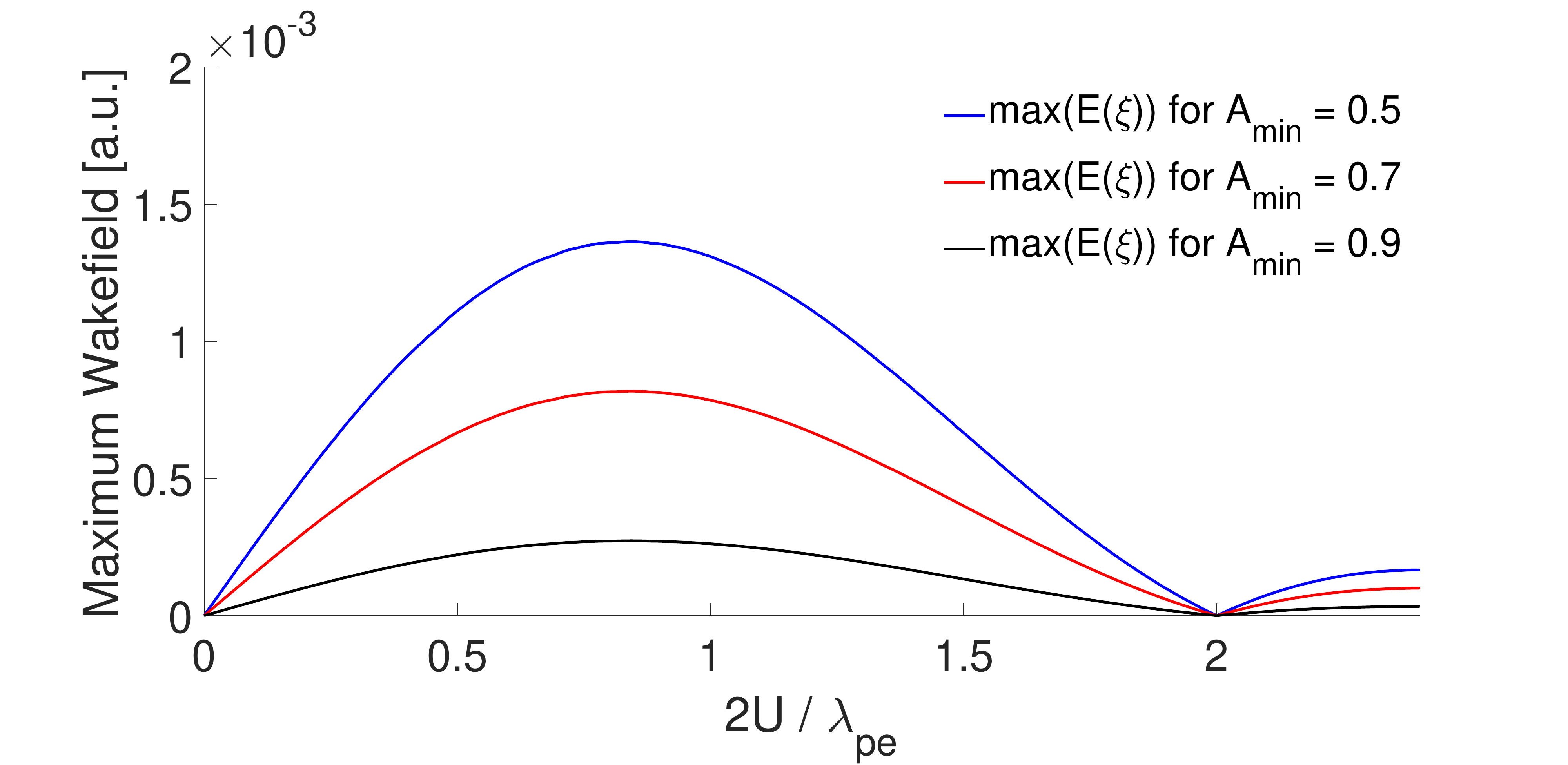}
	\caption{Maximum wakefields for different ratios of $2U/\lambda_{pe}$.}
	\label{fig: gap width}
\end{figure}

It can be easily seen from Eqn.\,\ref{eq: E1par} - \ref{eq: E3perp} that the amplitudes of the wakefields in the trailing part of the bunch are scaling linearly with the electron bunch charge $Q_{e}$, i.e. the depth of the gap in the effective charge density $A_0-A_{min}$.\\
Furthermore, the wakefields are strongly depending on the width of the gap in the effective charge density $2U$, corresponding to the longitudinal size of the injected electron bunch $\sigma_{z,e}$. The amplitude of the wakefields can be maximized for a given configuration of $A_{min}$ and $L$ and follow the distribution shown in Fig.\,\ref{fig: gap width}. For an electron bunch length $U$ on the order of $\lambda_{pe}$, the peak wakefields are the highest (see Fig.\,\ref{fig: gap width}).\\

\section{Discussion of the Method}
\begin{figure}[htb!]
	\begin{subfigure}[b]{\textwidth}
	\includegraphics[width=0.5\linewidth]{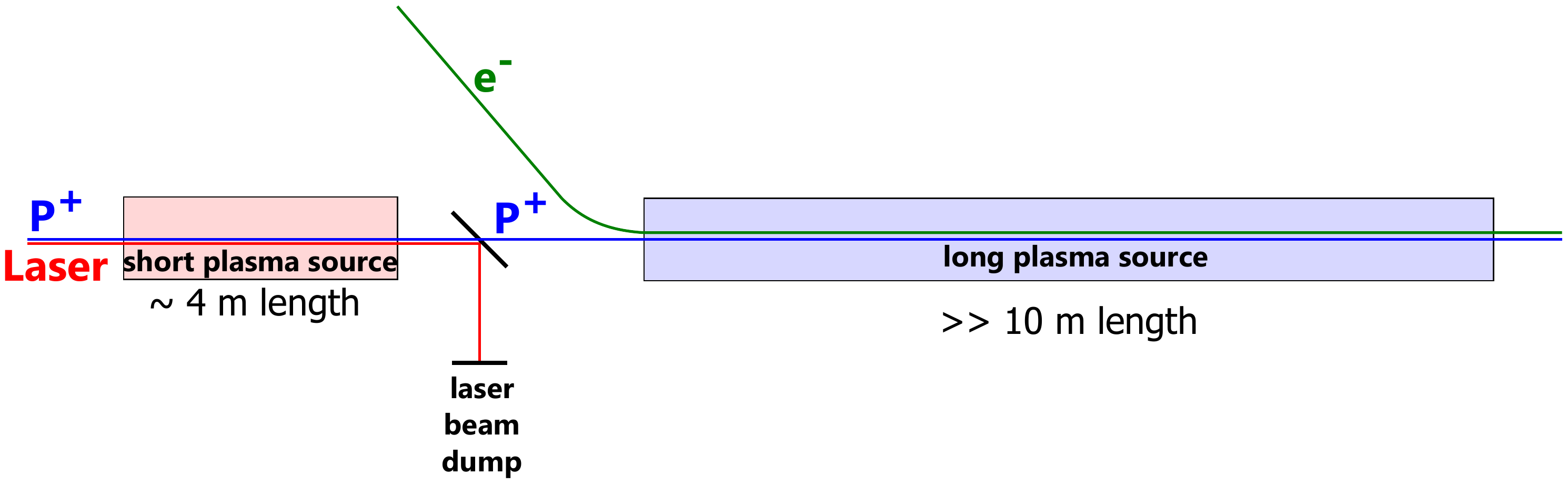}
	\caption{~~~~~~~~~~~~~~~~~~~~~~~~~~~~~}
	\label{fig:Run2-withLaser}
	\end{subfigure}
\begin{subfigure}[b]{\textwidth}
	\includegraphics[width=0.5\linewidth]{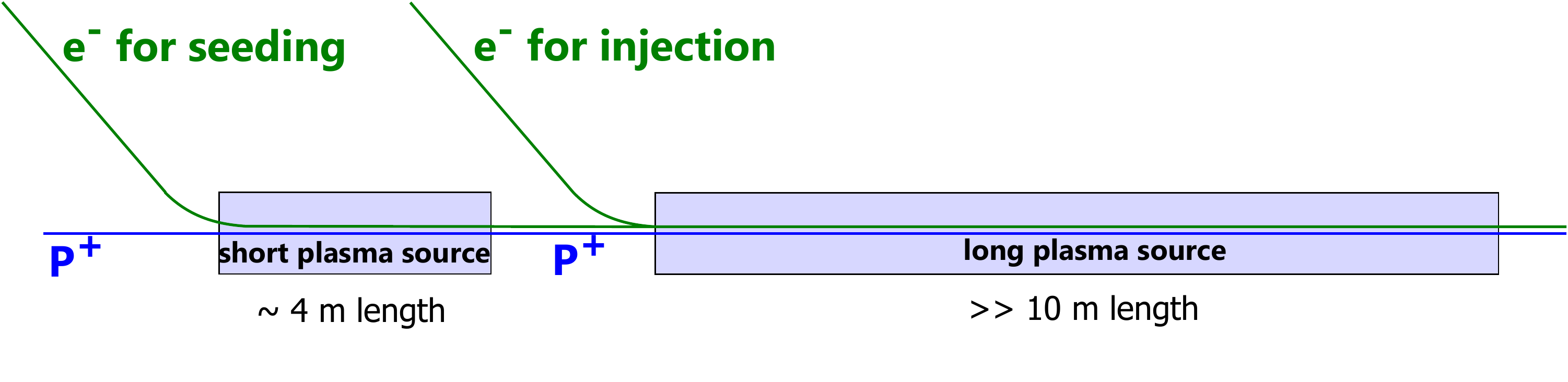}
	\caption{~~~~~~~~~~~~~~~~~~~~~~~~~~~~~}
	\label{fig:Run2-woLaser}
\end{subfigure}
\caption{Different schemes for a possible design of the split plasma cells in AWAKE Run 2. (a) Design with laser seeding and (b) design with seeding by electron injection.}
\label{fig:Run2-schemes}
\end{figure}

The seeding method described above could be an interesting approach for a seeding concept for the planned extension of the AWAKE experiment after 2021, called AWAKE Run 2 \cite{adli2016towards}. Current design studies for the plasma accelerator consist of a split plasma, a short one ($\sim$ 4\,m) for the seeding of the SSM and a second, longer one ($\sim$ 10\,m) driven by the modulated proton bunch for the acceleration of the electrons (see Fig.\, \ref{fig:Run2-schemes}). The injection point for the witness electron bunch is foreseen to be between the two plasmas. The short plasma source is similar to the one used in AWAKE Run 1, whereas the long plasma cell could be either a Helicon or a discharge source \cite{awake2016awake}.\\
By seeding the SMI with an electron bunch, as described here, the ionisation front is not necessary for seeding and thus the first plasma could also be preformed in one of these two types of sources. Hence, there would be no longer need for a maintenance-intensive high-power laser system nor for the complicated Rubidium handling and storage procedures required by a Rubidium based vapour source.\\
The major uncertainty concerning the feasibility of this approach is the effect of the first part of the proton bunch. Even though, according to the simulations of Kumar et. al. \cite{kumar2010self} and according to the wakefields described in Eq.\,\ref{eq: parallel and perpendicular wakefields}, there is no strong seeding of the SMI expected by the leading part of the bunch, the weak amplitude wakefields or those from noise may grow. Subsequently, wakefields in the front of the bunch would interfere with the seeded wakefields in the trailing part of the bunch. Hence, a phase-stable seeding of the SSM would no longer be possible.\\
First preliminary experimental results from the AWAKE-experiment show some evidence for a growth of wakefields along the bunch on a scale of a few meters as well as micro-bunching even without seeding, i.e. propagation in a preformed plasma.\\
In case the unseeded SM-growth in the leading part of the proton bunch is too high for a seeding with an electron bunch in the centre of the beam, another option would be to shift the electron bunch to an earlier position with respect to the proton bunch or even ahead of it (see Fig.\,\ref{fig:ElectronsAhead_SeedingScheme}). With this, the full proton bunch charge could be used for driving wakefields, while removing the risk of growth of SMI ahead of the seeding point. 
\section{Conclusions}
\label{sec:Conclusions}
We presented an alternative concept for the seeding of the SSM in plasma-wakefield accelerators. Instead of seeding by an ionisation front created by a high-power laser pulse, the seeding is archived by the injection of a short electron bunch in the centre of a long proton bunch propagating through a preformed plasma. Calculations in 2D linear theory show that the resulting steep rising edge in the effective charge density of the proton bunch drives large amplitude wakefields that seed the SSM. Although the approach is very promising and could simplify the design for AWAKE Run 2, first experimental results from the latest AWAKE Runs give evidence that wakefields do grow from noise in the front of the proton bunch, when the bunch propagates in a preformed plasma and no seeding is provided.

\appendix
\section{References}


\begin{thebibliography}{10}
	\expandafter\ifx\csname url\endcsname\relax
	\def\url#1{\texttt{#1}}\fi
	\expandafter\ifx\csname urlprefix\endcsname\relax\def\urlprefix{URL }\fi
	\expandafter\ifx\csname href\endcsname\relax
	\def\href#1#2{#2} \def\path#1{#1}\fi
	
	\bibitem{muggli2013physics}
	P.~Muggli, et~al., Physics of the {AWAKE} project, in: IPAC 2013: Proceedings
	of the 4th International Particle Accelerator Conference, 2013, pp.
	1179--1181.
	
	\bibitem{gschwendtner2016awake}
	E.~Gschwendtner, et~al., {AWAKE}, {T}he {A}dvanced {P}roton {D}riven {P}lasma
	{W}akefield {A}cceleration {E}xperiment at {CERN}, Nuclear Instruments and
	Methods in Physics Research Section A: Accelerators, Spectrometers, Detectors
	and Associated Equipment 829 (2016) 76--82.
	
	\bibitem{PhysRevLett.112.194801}
	K.~V. Lotov, A.~P. Sosedkin, A.~V. Petrenko, Long-term evolution of broken
	wakefields in finite-radius plasmas, Phys. Rev. Lett. 112 (2014) 194801.
	\newblock \href {http://dx.doi.org/10.1103/PhysRevLett.112.194801}
	{\path{doi:10.1103/PhysRevLett.112.194801}}.
	
	\bibitem{caldwell2016path}
	A.~Caldwell, et~al., Path to {AWAKE}: Evolution of the concept, Nuclear
	Instruments and Methods in Physics Research Section A: Accelerators,
	Spectrometers, Detectors and Associated Equipment 829 (2016) 3--16.
	
	\bibitem{kumar2010self}
	N.~Kumar, A.~Pukhov, K.~Lotov, Self-modulation instability of a long proton
	bunch in plasmas, Physical Review Letters 104~(25) (2010) 255003.
	
	\bibitem{lotov2013natural}
	K.~Lotov, et~al., Natural noise and external wakefield seeding in a
	proton-driven plasma accelerator, Physical Review Special Topics-Accelerators
	and Beams 16~(4) (2013) 041301.
	
	\bibitem{savard2017effect}
	N.~Savard, P.~Muggli, J.~Vieira, Effect of proton bunch parameter variation on
	awake, in: North American Particle Accelerator Conf.(NAPAC'16), Chicago, IL,
	USA, October 9-14, 2016, JACOW, Geneva, Switzerland, 2017, pp. 684--686.
	
	\bibitem{chen1987plasma}
	P.~Chen, et~al., Plasma focusing for high-energy beams, IEEE transactions on
	plasma science 15~(2) (1987) 218--225.
	
	\bibitem{adli2016towards}
	E.~Adli, Towards {AWAKE} applications: Electron beam acceleration in a proton
	driven plasma wake, in: 7th Int. Particle Accelerator Conf.(IPAC'16), Busan,
	Korea, May 8-13, 2016, JACOW, Geneva, Switzerland, 2016, pp. 2557--2560.
	
	\bibitem{awake2016awake}
	{AWAKE Collaboration}, {AWAKE} {S}tatus {R}eport 2016, SPSC-SR-194.
	
\end{thebibliography}

\end{document}